\documentclass[aps,prl,preprint,superscriptaddress,showpacs]{revtex4}

\usepackage{graphicx}
\usepackage{amsmath}
\usepackage{amssymb}

\begin{document}

\title{Mechanism of Collisional Spin Relaxation in $^3\Sigma$ Molecules}

\author{Wesley C. Campbell}
\email[]{wes@cua.harvard.edu}
\affiliation{Department of Physics, Harvard University, Cambridge,
  Massachusetts 02138, USA}
\affiliation{Harvard-MIT Center for Ultracold Atoms, Cambridge,
  Massachusetts 02138, USA}
\author{Timur V. Tscherbul}
\affiliation{Department of Chemistry, University of British Columbia,
  Vancouver, British Columbia V6T 1Z1, Canada}
\author{Hsin-I Lu}
\affiliation{School of Engineering and Applied Sciences, Harvard
  University, Cambridge, MA 02138 USA}
\affiliation{Harvard-MIT Center for Ultracold Atoms, Cambridge,
  Massachusetts 02138, USA}
\author{Edem Tsikata}
\affiliation{Department of Physics, Harvard University, Cambridge,
  Massachusetts 02138, USA}
\affiliation{Harvard-MIT Center for Ultracold Atoms, Cambridge,
  Massachusetts 02138, USA}
\author{Roman V. Krems}
\affiliation{Department of Chemistry, University of British Columbia,
  Vancouver, British Columbia V6T 1Z1, Canada}
\author{John M. Doyle}
\affiliation{Harvard-MIT Center for Ultracold Atoms, Cambridge,
  Massachusetts 02138, USA}
\affiliation{Department of Physics, Harvard University, Cambridge,
  Massachusetts 02138, USA}

\date{\today}

\begin{abstract}
We measure and theoretically determine the effect of molecular
rotational splitting on Zeeman relaxation rates in collisions of cold
$^3\Sigma$ molecules with helium atoms in a magnetic field.  All four
stable isotopomers of the imidogen (NH) molecule are magnetically
trapped and studied in collisions with $^3$He and $^4$He.  The $^4$He
data support the predicted $1/B_e^2$ dependence of the
collision-induced Zeeman relaxation rate coefficient on the molecular
rotational constant $B_e$.  The measured $^3$He rate coefficients are
much larger than $^4$He and depend less strongly on $B_e$, and the
theoretical analysis indicates they are strongly affected by a shape
resonance.  The results demonstrate the influence of molecular
structure on collisional energy transfer at low temperatures.

\end{abstract}


\maketitle

The development of experimental techniques for cooling molecular
ensembles to subKelvin temperatures has opened up possibilities to
study collision dynamics of molecules in a new, previously
inaccessible, regime.  Elastic and inelastic scattering of molecules
at such low temperatures is sensitive to external electromagnetic
fields as well as fine and hyperfine interactions, of negligible
importance in thermal collisions.  Cold collisions hold the promise of
being exploited to study a new class of processes including chemical
reactions induced by fine non-adiabatic interactions
\cite{KremsPCCP08}, field control of intermolecular interactions
\cite{KremsIRPC05}, effect of the geometric phase on collision
dynamics \cite{WredeSCIENCE05} and the effect of long-range
intermolecular interactions in determining chemical reactions
\cite{BohnPRA05}.  Cooling techniques also allow for the preparation
of molecules in a single quantum state and enable experimental tests
of fundamental theories of molecular dynamics.  Of particular
importance for the experimental work on thermal isolation of molecular
ensembles in magnetic traps and collisional cooling of molecules to
ultracold temperatures is the theory of collision-induced Zeeman
relaxation \cite{DalgarnoJCP04}.  Quantum mechanical calculations of
Krems and Dalgarno showed that the rates of Zeeman relaxation in
collisions of open-shell molecules in $\Sigma$ electronic states with
cold atoms are very sensitive to the rotational constant of the
molecules, which can be used to identify the range of molecules
amenable to dissipative cooling in magnetic traps based on their
rotational structure.  In this Letter, we present the first
experimental study of the effect of the rotational constant on the
magnetic Zeeman relaxation.

Zeeman transitions in $\Sigma$-state molecules are thought to be
induced by second-order interactions of the electron spin with the
rotational angular momentum \cite{DalgarnoJCP04}.  The same
interactions give rise to an intricate interplay of the fine structure
couplings and the couplings induced by electrostatic interaction
forces between molecules, which can be used to engineer novel quantum
phases with molecules trapped on optical lattices \cite{DemlerPRL06}
or develop schemes for quantum information processing based on
state-dependent interactions between molecules \cite{DeMillePRL02}.
It is therefore extremely important to validate by experimental
measurements the theory of these interactions subject to uncertainties
due to inaccuracies of quantum chemistry calculations.  In the present
work, we study collisions of magnetically trapped imidogen (NH)
molecules with helium atoms.  Our experimental data not only exemplify
the effects of internal molecular structure on collision-induced
Zeeman relaxation of $^3\Sigma$ molecules, but also provide evidence
for the role of scattering resonances in a multiple-partial wave
collision regime and provide benchmark results for tests of ab initio
theories of molecular dynamics at cold temperatures.

In this work, we vary the rotational constant of the molecule ($B_e$)
by changing the NH isotopomers and the reduced mass of the collision
system by changing the isotope of He.  We present the measurement data
for all eight isotopic combinations of the imidogen-helium collision
system ($^x$N$^y$H--$^z$He).  The predicted $1/B_e^2$ dependence of
the inelastic cross section \cite{KremsJCP05} fits well for the $^4$He
data, supporting the model that the couplings between rotational
states of the molecule drive helium induced Zeeman relaxation.  The
measured $^3$He inelastic collision rate coefficient is found to be
much larger than that for $^4$He, and the $^3$He data show a weaker
dependence on the molecular rotational splitting.  Based on quantum
mechanical calculations, we show that the anomalous behavior of the
$^3$He-induced Zeeman relaxation rate coefficient of imidogen can be
attributed to the presence of a predicted shape resonance
\cite{DoylePRL07, KremsJCP05}.

Our experimental apparatus is described in detail elsewhere
\cite{DoylePRL07}.  A molecular beam of imidogen radicals is loaded
into a cryogenic buffer-gas cell through a 1 cm diameter molecular
beam entrance aperture.  The buffer-gas cell is thermally connected to
a $^3$He refrigerator through a copper heat link, and resides in the
bore of a superconducting anti-Helmholtz magnet.  The trap magnet
creates a spherical quadrupole field up to 3.9 T deep centered inside
the buffer-gas cell.  Molecules entering the cell collide with the
cold helium buffer gas and the low-field seeking (LFS) molecules fall
into the magnetic trap.  The buffer gas is continuously supplied to
the cell by a fill line that is thermally connected to the $^3$He
refrigerator.  The buffer-gas density is set by the flow rate and the
conductance out of the molecular beam input aperture.  The buffer-gas
density is monitored and controlled by a flow controller at room
temperature, and the buffer-gas density has been calibrated previously
\cite{DoylePRL07}.

Trapped radicals are detected using laser-induced fluorescence (LIF).
A cw dye laser is frequency doubled to provide 336 nm excitation light for
imidogen.  The beam enters and exits the cell through windows in the
side, and fluorescence is also collected through a window in the side
of the cell perpendicular to the excitation beam.  Fluorescence from
the trap region is imaged onto the face of a photomultiplier tube
operating in photon-counting mode.  In this work, detection takes
place via LIF excited on the $A^3\Pi_2(v=0,J=2) \leftarrow
X^3\Sigma^-(v^{\prime\prime}=0,J^{\prime\prime}=1)$ transition from
the LFS Zeeman sublevel of the rotational ground state.

The molecular beam of radicals is produced in a DC glow discharge of
ammonia \cite{DoyleEPJD04} and each imidogen isotopomer is created by
using a different isotopic variant of ammonia as the stagnation gas
($^{14}\mathrm{NH}_3$ for $^{14}\mathrm{NH}$, $^{15}\mathrm{ND}_3$ for
$^{15}\mathrm{ND}$, etc.).  Isotopic purity levels for all four gases
are higher than 95\% and we are unable to detect impurity
contributions in the LIF signal.  Buffer gas isotopes are interchanged
by feeding from different bottles, and the helium isotopic purity is
99.95\% or better.  In the experiments with $^4$He, the temperature of
the cell is kept above 700 mK to prevent saturation of the vapor
pressure.  The isotope shifts for all four stable isotopomers of
imidogen make spectroscopic distinction between them possible, even in
the presence of the magnetic trapping field.  The observed shifts of
the $A^3\Pi_2(v=0,N=1,J=2) \leftarrow X^3\Sigma^-(v^{\prime\prime}=0,
N^{\prime\prime}=0, J^{\prime\prime}=1)$ transition from the dominant
isotopomer ($^{14}$NH) are $+11.70 \mbox{ cm}^{-1}$ ($^{14}$ND),
$+0.12 \mbox{ cm}^{-1}$ ($^{15}$NH), and $+11.83 \mbox{ cm}^{-1}$
($^{15}$ND).

\begin{figure}
\includegraphics[width=3in]{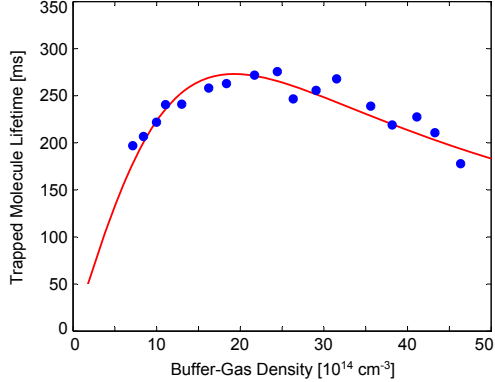}
\caption{Measured trap lifetime vs. buffer-gas density for $^{14}$NH
  with $^4$He at 740 mK.  The solid curve is fit to extract a
  collision-induced Zeeman relaxation rate coefficient.\label{K1}}
\end{figure}

Figure \ref{K1} shows the trap lifetime of $^{14}$NH as a function of
$^4$He density at 740 mK.  The lifetime is measured by fitting a
single-exponential to the fluorescence signal \cite{DoylePRL07}.  In
the low-density regime on the left side, the trap lifetime increases
with buffer-gas density.  This is because the buffer gas enforces
diffusive motion of the molecules, effectively making them slower to
exit the trap.  The lifetime then reaches a maximum as
collision-induced Zeeman relaxation becomes more frequent and
decreases as $1/n$ for high buffer-gas densities ($n$).  The solid
curve is a two-parameter fit to the data of the form
$1/\tau_{\mathrm{eff}} = A/n + nk_{\mathrm{ZR}}$ where
$\tau_{\mathrm{eff}}$ is the trap lifetime, $A$ is a fitting parameter
describing the influence of elastic collisions and trap depth on the
lifetime, and $k_{\mathrm{ZR}}$ is the collision-induced Zeeman
relaxation rate coefficient.  We measure $k_{\mathrm{ZR}}$ by fitting
the trap lifetime as a function of buffer-gas density for each of the
eight imidogen-helium collision pairs.

\begin{table}
\caption{Zeeman relaxation rate coefficients for imidogen in
  collisions with helium in units of $10^{-15} \mbox{
  cm}^3\mbox{s}^{-1}$.  The quoted uncertainties are statistical. The
  uncertainty in the absolute buffer-gas density results in a
  systematic uncertainty of $\pm 30\%$ for all measurements.  The last
  two columns show the results of our calculations and the values in
  parentheses are calculated with the interaction anisotropy
  multiplied by 1.6}
\begin{tabular}{|r|c|c|   c|c| }
\hline
Collision Species & $^3$He\footnote{Experiment: $T = 580 - 633 \mbox{ mK}$} & $^4$He\footnote{Experiment: $T = 720 - 741 \mbox{ mK}$} 
&   $^3$He\footnote{Theory: $T = 600 \mbox{ mK}$} 
 &   $^4$He\footnote{Theory: $T = 700 \mbox{ mK}$}
\\
\hline
$^{14}$NH & $4.5 \pm 0.3$& $1.1 \pm 0.1$  &
$0.56$ $(2.1)$  &  $0.23$ $(0.76)$ 
\\  
\hline
$^{15}$NH & $5.1 \pm 0.4$& $1.4 \pm 0.2$  &
$0.61$ $(2.3)$   &    $0.27$ $(0.87)$
\\
\hline
$^{14}$ND & $9.3 \pm 0.8$& $4.0 \pm 0.7$  &
$2.17$  $(9.1)$   &   $1.03$  $(3.9)$  
\\
\hline
$^{15}$ND & $13.0 \pm 0.8$& $2.8 \pm 0.6$ &
$2.32$ $(9.1)$   & $1.18$ $(4.3)$
\\
\hline
\end{tabular}
\label{KTable}
\end{table}

The results of the measurements are summarized in Table \ref{KTable}
and are best interpreted in the context of theory put forth in
Ref. \cite{DalgarnoJCP04}.  Krems and Dalgarno demonstrated that
collision-induced spin-depolarization in $^3\Sigma$ molecules is
mediated by a small admixture of the anisotropic rotational state
$|N=2\rangle$ in the rotational ground state (nominally $|N=0\rangle$)
of the molecule due to the spin-spin interaction.  The electrostatic
interaction with helium ($V_{\mathrm{He}}$) cannot directly couple
different Zeeman sublevels of an $|N=0\rangle$ rotational state, but
there can be a nonzero off-diagonal contribution $\langle N=2 \left|
V_{\mathrm{He}} \right| N=0 \rangle$ to the Zeeman transition
probability for $^3\Sigma$ molecules due to the $N=2$ contribution.
The admixture of $|N=2\rangle$ in the ground rotational state is
determined to first order by the ratio $\lambda_{\mathrm{SS}}/B_e$,
where $\lambda_{\mathrm{SS}}$ is the spin-spin interaction constant.
The collision-induced Zeeman relaxation cross section is then
predicted to scale as $\lambda_{\mathrm{SS}}^2/B_e^2$
\cite{KremsJCP05}.  This is in direct contrast with $^2\Sigma$
molecules, where the collision-induced Zeeman transition is induced by
the spin-rotation interaction in the molecule, and is predicted to
scale as $\gamma_{\mathrm{SR}}^2/B_e^4$.

\begin{figure}
\includegraphics[width=3in]{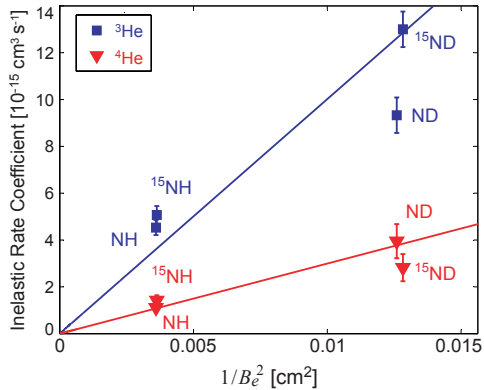}
\caption{Collision-induced Zeeman relaxation (inelastic collision)
  rate coefficient as a function of the rotational constant ($B_e$) of
  the imidogen radical.  The solid lines are one-parameter fits to a
  $1/B_e^2$ scaling law for the inelastic collision rate coefficients.
  The rotational constants for $^{15}$N-bearing isotopomers are
  estimated from the $^{14}$N-bearing rotational constants and the
  reduced masses: $B_e^{\prime} = B_e \times
  \mu/\mu^{\prime}$.\label{KvsRot}}
\end{figure}

To show the dependence of the rate coefficients on $B_e$, we plot
$k_{\mathrm{ZR}}$ extracted from our data vs $1/B_e^2$ in
Fig. \ref{KvsRot}.  The solid curves are one-parameter fits to a
$B_e^{-2}$ scaling of the inelastic rate coefficient.  The measured
data for $^4$He are in good agreement with the scaling prediction.
While the scaling for $^3$He does not follow $1/B_e^2$, in all cases,
decreasing the imidogen rotational constant by a factor of $\approx 2$
(through H$\rightarrow$D substitution) still dramatically increases
the Zeeman relaxation rate.

In order to fully understand the experimental observations and their
implications, we computed the Zeeman relaxation rate coefficients for
each pair of the collision partners using a rigorous quantum
scattering approach described in Ref. \cite{KremsPRA03}. The
calculations are based on the most accurate NH--He interaction
potential borrowed from Ref. \cite{KremsJCP05}. The rate coefficients
in the temperature interval $T=0.2-1$ K were obtained from the cross
sections calculated at 300 collision energies from 0.01 to 1.5
cm$^{-1}$ and summed over all final spin states of NH.  Table
\ref{KTable} shows that the computed rate coefficients for Zeeman
relaxation are smaller than the measured values by a factor between
2.3 and 8. At the same time, the computed cross section for elastic
collisions of $^{14}$NH with $^3$He is in excellent agreement with the
experimental measurement ({\it cf.,} Refs.
\cite{DoylePRL07,KremsJCP05}). This indicates that the isotropic part
of the He--NH interaction potential from Ref. \cite{KremsJCP05} is
accurate but the anisotropy of the atom--molecule potential may be
significantly underestimated. In order to elucidate the effects of the
interaction anisotropy on the Zeeman relaxation process, we
re-computed the rate coefficients with the interaction potential
anisotropy multiplied by 1.6. The $60$\% change of the interaction
anisotropy enhances the rate coefficients for Zeeman relaxation to a
great extent but leaves the rate coefficients for elastic collisions
almost unaffected.  The rate coefficients calculated with the modified
potential agree to within $30 \%$ with the experimental data for all
collision systems except $^{14}$NH--$^3$He and $^{15}$NH--$^3$He. The
measured effect of the NH/ND isotopic substitution on collisions with
$^3$He is thus again in marked disagreement with theory.

\begin{figure}
\includegraphics[width=3in]{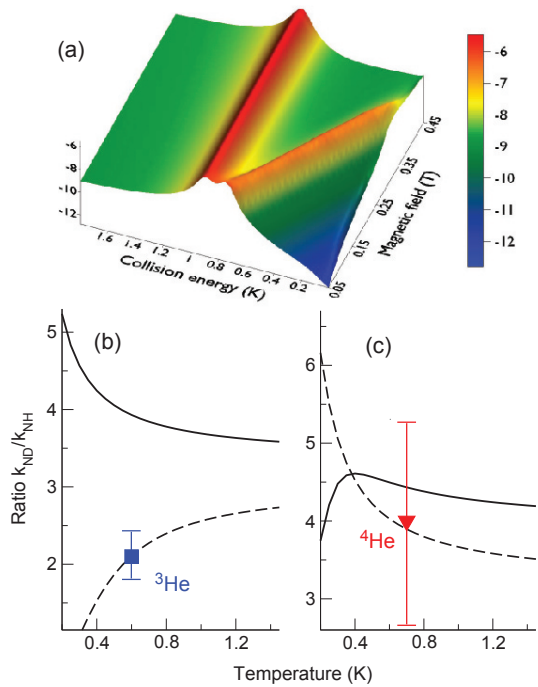}
\caption{Upper panel (a): The logarithm of the cross section for the
$|M_S=1\rangle \to |M_S=-1\rangle$ transition in collisions with
$^3$He atoms as a function of the magnetic field and collision energy.
Lower panel: The calculated ratios of the Zeeman relaxation rate
coefficients $k_\text{ND}/k_\text{NH}$ for $^3$He (b) and $^4$He (c)
obtained with the original NH--He interaction potential (full lines)
and with the potential multiplied by 1.1 (dashed lines). The
experimental data are shown as symbols. \label{TheoryRatio}}
\end{figure}

Table \ref{KTable} demonstrates that the rates of Zeeman relaxation in
collisions of NH and ND molecules with $^3$He, both measured and
calculated, are consistently larger than in collisions with
$^4$He. The effect of nuclear spin of $^3$He should be negligible
\cite{JonathanThesis}.  This enhancement suggests the presence of
scattering resonances. Figure \ref{TheoryRatio}(a) shows the cross
section for the dominant $|M_S=1\rangle \to |M_S=-1\rangle$ transition
in the ground rotational state of NH induced by collisions with
$^{3}$He atoms computed as a function of the incident collision energy
and the magnetic field magnitude. The cross section displays a single
resonance peak that splits into two at finite magnetic field. The
higher-energy resonance is insensitive to the magnetic field, which
suggests that it is a shape resonance in the incoming collision
channel. The low-energy resonance can be classified as a shape
resonance in the outgoing collision channel \cite{ZygelmanJPB02}. The
$|M_S=-1\rangle$ state is $4\mu_0B$ lower in energy than the initial
$|M_S=1\rangle$ state, where $\mu_0$ is the Bohr magneton, and the
splitting between the resonances in Fig. \ref{TheoryRatio}(a)
increases linearly with the magnetic field strength. The cross
sections for collisions of ND and $^3$He display a similar resonance
pattern.  However, the cross sections calculated for Zeeman relaxation
in collisions of NH and ND molecules with $^4$He show no resonance
structures in the energy interval $0.1$--$1.5$ cm$^{-1}$ corresponding
to the temperature of the experimental measurements.

The shape resonances affect the temperature dependence of the Zeeman
relaxation rates and modify the scaling of the rates with $B_e$ (see
Fig. \ref{TheoryRatio}). To verify this, we re-computed the rate
constants for Zeeman relaxation with the interaction potential
multiplied by 1.1.  Figures \ref{TheoryRatio}(b) and (c) display the
ratios of the spin relaxation rates $k_\text{ND}/k_\text{NH}$ for ND
and NH at a magnetic field of 0.1 T computed both with the original
and modified potentials.  The calculated ratios for collisions with
$^4$He are in excellent agreement with the experimental data and the
$1/B_e^2$ scaling law.  We focus here on the $^{14}$N data, but Table
I shows that the predicted rates for $^{15}$N are similar.  For
$^3$He, the results obtained with the original potential overestimate
the measured $k_\text{ND}/k_\text{NH}$ ratio by a factor of 2 ({\it
cf.,} Table I) but the results obtained with the modified potential
agree well with the measurements. Multiplying the potential by 1.1
moves the resonance structure to lower energies and changes the
temperature dependence of the rate coefficients dramatically,
weakening the dependence on $B_e$.  The agreement between our data and
the scaled potential suggests that the imidogen-helium potential used
for our calculation is too shallow.

In conclusion, we have trapped both fermionic (NH, $^{15}$ND) and
bosonic (ND, $^{15}$NH) imidogen radicals in a magnetic trap using
buffer-gas loading.  We have measured the Zeeman relaxation rate
coefficients for all stable isotopes of NH in collision with both
stable isotopes of helium at $\approx 650 \mbox{ mK}$.  The
experiments show that the Zeeman relaxation rates for $^3\Sigma$
molecules in the ground rotational state are sensitive to the
rotational constant of the molecule.  The observed scaling behavior of
the collision-induced Zeeman relaxation rate, which is compatible with
the prediction of $1/B_e^2$, has important implications for
sympathetic cooling of magnetically trapped molecules, and suggests
that $^3\Sigma$ molecules with large rotational splitting will be more
stable against spin-projection changing collisions.  Furthermore, the
rotational constant of NH can be identified as a general marker in the
magnetic trapping landscape---$^3\Sigma$ molecules with smaller
rotational constants will generally be difficult to cool collisionally
in a magnetic trap.

Our study also underlines the role of multiple partial-wave scattering
in collision dynamics at cold temperatures.  The results of rigorous
quantum calculations demonstrate that the collision dynamics of NH and
ND molecules with $^3$He at the temperatures of our measurements are
modified by a scattering shape resonance.  The resonance enhances the
inelastic collision rate, alters the temperature dependence of the
collision rate coefficients and modifies the dependence on the
rotational constant.  Identification of resonances in candidate
collisional cooling systems will therefore be an important guide for
future experiments using sympathetic and evaporative cooling to reach
the ultracold regime.

\begin{acknowledgments}
This work was supported by the U.S. Department of Energy under
Contract No. DE-FG02-02ER15316, NSERC of Canada and the U.S. Army
Research Office.  T. T. was supported by Killam Trusts.
\end{acknowledgments}

\bibliography{Wesbib2.bib}

\end{document}